\documentclass[floats,prl,aps,showpacs,amssymb,nofootinbib,twocolumn]{revtex4}
\usepackage{amssymb}
\usepackage{graphics}
\usepackage{amsmath}
\usepackage{amsfonts}
\usepackage{bm}% bold math
\usepackage[]{latexsym}
\usepackage{float}

\newcommand{\be}{\begin{equation}}\newcommand{\ee}{\end{equation}}
\newcommand{\bea}{\begin{eqnarray}}\newcommand{\eea}{\end{eqnarray}}
\newcommand{\brr}{\begin{array}}\newcommand{\err}{\end{array}}
\newcommand{\bit}{\begin{itemize}}\newcommand{\eit}{\end{itemize}}
\newcommand{\ben}{\begin{enumerate}}\newcommand{\een}{\end{enumerate}}

\newcommand{\ba}{\begin{array}}
\newcommand{\ea}{\end{array}}

\def\lf{\left}

\def\non{\nonumber}

\def\ri{\right}
\def\al{\alpha}

\def\la{\lambda}

\def\1{{_{1}}}\def\2{{_{2}}}

\def\noHe0{:\;\!\!\;\!\!:H_e(0):\;\!\!\;\!\!:}
\def\noHm0{:\;\!\!\;\!\!:H_\mu(0):\;\!\!\;\!\!:}

\def\lf{\left}

\def\non{\nonumber}

\def\ri{\right}

\def\al{\alpha}

\def\la{\lambda}

\def\1{{_{1}}}\def\2{{_{2}}}

\begin{document}
\title{%General covariance, Unruh effect and neutrino oscillations
Neutrino oscillations in the Unruh radiation}

\author{Massimo Blasone\footnote{blasone@sa.infn.it}$^{\hspace{0.3mm}1,2}$, Gaetano Lambiase\footnote{lambiase@sa.infn.it}$^{\hspace{0.3mm}1,2}$, Giuseppe Gaetano Luciano\footnote{gluciano@sa.infn.it}$^{\hspace{0.3mm}2}$ 
and Luciano Petruzziello\footnote{lpetruzziello@na.infn.it}$^{\hspace{0.3mm}1,2}$} \affiliation
{$^1$Dipartimento di Fisica, Universit\`a di Salerno, Via Giovanni Paolo II, 132 I-84084 Fisciano (SA), Italy.\\ $^2$INFN, Sezione di Napoli, Gruppo collegato di Salerno, Italy.}

\date{\today}
\def\be{\begin{equation}}
\def\ee{\end{equation}}
\def\al{\alpha}
\def\bea{\begin{eqnarray}}
\def\eea{\end{eqnarray}}

\begin{abstract}
The study of the decay of an accelerated proton recently provided a ``theoretical proof'' of the Unruh effect. On the basis of general covariance of Quantum Field Theory, indeed, it was found that the decay rates in the inertial and comoving frames do coincide only when the thermal nature of the accelerated vacuum is taken into account. 
Such an analysis  was then extended to the case with neutrino mixing. In this Letter, we show that, by further embedding neutrino oscillations in the above framework, general covariance necessarily entails the use of flavor neutrinos as asymptotic states, as well as the occurrence of neutrino oscillations in the Unruh thermal bath.

\end{abstract}

\vskip -1.0 truecm 

\maketitle

{\em Introduction}. In the history of Physics, the adoption of principles has revealed to be a formidable investigation tool. 
Although intimately related to the phenomenological realm from which they stem,  once elevated to the status of postulates, physical principles act as lighthouses for the development of a consistent theoretical apparatus. Paradigmatic examples are the principle of conservation of energy, which led for instance to the discovery of the neutrino, and the principle of constancy of speed of light, at the basis of the construction of Special Relativity.
 
Recently, general covariance has been advocated to exhibit that the internal consistency of quantum field theory (QFT) unavoidably requires the existence of the Unruh effect, also known as Fulling-Davies-Unruh effect~\cite{Fulling,Davies,Unruh:1976db}. Indeed, starting from the statement that acceleration can influence even the proper lifetime of stable particles~\cite{Muller:1997rt}, in a series of remarkable papers~\cite{Matsas:1999jx} it was shown that the tree-level decay rate of an accelerated proton via inverse $\beta$-decay is frame-independent only when the thermal nature of the vacuum for a non-inertial observer is considered.

The fact that a theoretical requirement leads to the specific form of the ground state for an accelerated observer should be regarded as a considerable result, especially in view of the perplexities which have been sometimes raised about the physical significance of the Unruh 
effect~\cite{scettici}. Such a skepticism is enhanced by the lack of direct evidences of this phenomenon, as it also happens for the case of the Hawking radiation~\cite{Hawking}. In fact, at present the most likely arena for (indirect) experimental tests of these effects is given by analogue gravity~\cite{Barcelo}.

In the above cited studies on the accelerated proton decay, the emitted neutrino was initially treated as massless and only in Ref.~\cite{suzuki} as a massive particle.
In these  works, however, neutrino mixing was not taken into account. This was done for the first time in Ref.~\cite{Aluw}, where a discrepancy between the proton decay rate in the inertial and comoving frames was claimed to arise. Subsequently, it was proved that general covariance does indeed hold in the above analysis: this has been achieved by employing either flavor~\cite{Blasone:2018czm} or mass~\cite{Cozzella:2018qew} eigenstates for neutrinos, thus leaving an essential ambiguity on the very nature of asymptotic neutrino states. 

In this Letter we show that, due to the occurrence of neutrino oscillations in the problem at hand, general covariance  leads to the conclusion that flavor eigenstates are   fundamental and neutrinos belonging to the Unruh radiation must oscillate. Throughout the paper, we use natural units $\hbar=c=~1$  and the Minkowski metric with the mostly negative signature.

Let us start by setting the framework. Following the approach and the notation of Refs.~\cite{Matsas:1999jx}, the proton $|p\rangle$ and the neutron $|n\rangle$ can be viewed as unexcited and excited states of the nucleon, respectively. In addition, we assume to deal with particles that are energetic enough to possess a well-defined trajectory. In these conditions, it is possible to employ the Fermi theory of current-current interaction, where we consider a quantum leptonic and a classical hadronic current $\hat{J}^{\mu}_{\ell}\hat{J}_{h,\mu}~\hspace{-1mm}\rightarrow~\hspace{-1mm}\hat{J}^{\mu}_{\ell}\hat{J}^{(cl)}_{h,\mu}$, with
$\hat{J}^{(cl)}_{h,\mu}=\hat{q}(\tau)u_{\mu}\delta(x)\delta(y)\delta(u-a^{-1})$.
Here\footnote{We assume the acceleration to occur along the $z$-direction. With this choice, the Rindler coordinates $(v, x, y, u)$  are related with the Minkowski coordinates $(t, x, y, z)$ as follows: $t=u\sinh{v}$, $z=u\cosh{v}$, with $x$ and $y$ untouched.}, $\tau=v/a$ is the nucleon's proper time (with $v$ being the Rindler time coordinate), $a$ its proper acceleration and $u~\!=~\!a^{-1}~\!=\!~\mathrm{const}$ represents the spatial Rindler coordinate which denotes the world line of the uniformly accelerated particle. The four-velocity of the nucleon $u^{\mu}$ is given by
$
u^{\mu}=(a, 0, 0, 0), \, u^{\mu}=(\sqrt{a^2t^2+1}, 0, 0, at)$,
in Rindler and Minkowski coordinates, respectively. 
In accordance with Refs.~\cite{Matsas:1999jx, Birrell}, the Hermitian monopole $\hat{q}(\tau)$  is given by
$
\hat{q}(\tau)\equiv e^{i\hat{H}\tau}\hat {q}_0e^{-i\hat{H}\tau},
$ 
where $\hat{H}$ is the nucleon Hamiltonian and $\hat {q}_0$ is used to reconstruct the effective Fermi constant
$
G_F\equiv\langle p|\hspace{0.2mm}\hat q_0|n\rangle.
$

Assuming to deal with a simplified two-flavor model, the interaction of the charged leptons $\hat{\Psi}_{\alpha}$ and neutrinos $\hat\Psi_{\nu_\alpha}$ $(\alpha=e,\mu)$ with the nucleon current $\hat{J}^{(cl)}_{h,\mu}$ is described by the Fermi action
\be
\label{eqn:Fermiaction}
 \hat{S}_{I}\! =\!\sum_{\alpha=e,\mu}\hspace{-0.5mm}\int d^{4}x\sqrt{-g}\hat{J}^{(cl)}_{h,\la}\hspace{-0.5mm}\left(\hat{\overline{\Psi}}_{\nu_\alpha}\gamma^{\la}\hat{\Psi}_{\alpha} + \hat{\overline{\Psi}}_{\alpha}\gamma^{\la}\hat{\Psi}_{\nu_\alpha}\right),\!
\ee
where $g\equiv \mathrm{det}(g_{\mu\nu})$ and $\gamma^\la$ are the gamma matrices in Dirac representation (e.g., see Ref.~\cite{Itzykson}).
In Eq.~\eqref{eqn:Fermiaction}, neutrino fields 
with definite flavors are related to the ones with
definite masses by
\begin{subequations}\label{Pontecfield}
\begin{align}
\label{p1field}
\hat{\Psi}_{\nu_e}&=\mathrm{cos}\,\theta\,\hat{\Psi}_{\nu_1}+\mathrm{sin}\,\theta\,\hat{\Psi}_{\nu_2},\\[1mm]
\label{p2field}
\hat{\Psi}_{\nu_\mu}&=-\mathrm{sin}\,\theta\,\hat{\Psi}_{\nu_1}+\mathrm{cos}\,\theta\,\hat{\Psi}_{\nu_2}.
\end{align}
\end{subequations}
In what follows, however, we shall focus on the processes 
involving only the production of a positron. The case in which an anti-muon appears in the final state can be treated analogously and in an independent way.
%, thus preventing the generality of the upcoming discussions from being affected by this assumption.

\medskip

%\subsection*{Inertial frame}

{\em Inertial frame}. In the inertial frame, the process we consider is the decay of an accelerated proton via inverse $\beta$-decay
\be\label{i}
\mathrm{(i)}\quad p\,\to\, n\, +\, e^{+}\,+\,\nu_e\,,
\ee
which is pictorially represented in Fig.~1.

The fermion field  is expanded as
\be
\label{inertexp}
\hspace{-.5mm}\hat{\Psi}\!=\!\sum_{\sigma=\pm}\!\int\!\!\frac{d^3k}{4\pi^{\frac{3}{2}}}\left[{e^{-ik^\mu x_\mu}} u_{\sigma}^{(\omega)}\hat{b}_{\textbf{k}\sigma}+{e^{ik^\mu x_\mu}} u_{-\sigma}^{(-\omega)}\hat{d}_{\bf{k}\sigma}^{\dagger}\right]\!,\!\!
\ee
where $\sigma$ is the polarization,  $\omega=\sqrt{k^2+m^2}$ and $u_{\sigma}^{(\omega)}$ is the spinor defined as in Ref.~\cite{Blasone:2018czm}.

The tree-level transition amplitude for the process (i) reads~\cite{Blasone:2018czm}
\bea\label{tramp}
&\mathcal{A}^{(\nu_e)}_{\mathrm{(i)}}&=\langle n|\otimes\langle e^{+},\nu_{e}|\hat{S}_{I}|0\rangle\otimes|p\rangle\\[2mm]\nonumber
&=&\hspace{-3.5mm}\frac{G_F}{2^4\pi^3}\Big[\cos^2\theta\, \mathcal{I}_{\sigma_\nu\sigma_e}(\omega_{\nu_1},\omega_e)\,+\, \sin^2\theta\, \mathcal{I}_{\sigma_\nu\sigma_e}(\omega_{\nu_2},\omega_e)\Big],
\eea
where for simplicity we have omitted the $\textbf{k}$- and $\sigma$-dependence of the lepton states, we have assumed equal momenta and polarizations for neutrinos with definite masses and
\bea
\label{I}
\hspace{-25mm}&\mathcal{I}_{\sigma_{\nu}\sigma_e}(\omega_{\nu_j}, \omega_e)& = \hspace{-0.5mm}\int_{-\infty}^{+\infty}\hspace{-2.3mm}d\tau\,u_{\mu}\left[\bar{u}_{\sigma_\nu}^{(+\omega_{\nu_j})}\gamma^{\mu}{u}_{-\sigma_e}^{(-\omega_e)}\right]\\[2mm]\nonumber
&\times&\hspace{-10mm}e^{i\big[\Delta m\hspace{0.2mm}\tau\,+\,a^{-1}\left(\omega_{\nu_j}\,+\,\omega_{e}\right)\sinh a\tau\,-\,a^{-1}\left(k_\nu^z\,+\,k^z_{e}\right)\cosh a\tau\big]}\,.
\eea
with $j=1,2$.

Note that the asymptotic flavor neutrino state $|\nu_e\rangle$
in Eq.~\eqref{tramp} has been expressed in terms
of the corresponding mass states $|\nu_i\rangle$ $(i=1,2)$ 
by means of Pontecorvo mixing transformations \cite{Bilenky:1978nj} 
\begin{subequations}\label{Ponte}
\begin{align}
\label{p1}
|\nu_e\rangle&=\mathrm{cos}\,\theta\,|\nu_1\rangle+\mathrm{sin}\,\theta\,|\nu_2\rangle,\\[2mm]
\label{p2}
|\nu_\mu\rangle&=-\mathrm{sin}\theta\,|\nu_1\rangle+\mathrm{cos}\,\theta\,|\nu_2\rangle.
\end{align}
\end{subequations}
The differential transition probability is defined as
\be
\label{drt}
\frac{d^{6}\mathcal{P}_{\mathrm{(i)}}^{(\nu_e)}}{d^3k_{\nu}\,d^3k_{e}}=\sum_{\sigma_e,\sigma_{\nu}}\lf|\mathcal{A}^{(\nu_e)}_{\mathrm{(i)}}\ri|^2,
\ee
and the transition rate is given by $\Gamma=\mathcal{P}/T$, with 
\be T\equiv \int_{-\infty}^{+\infty}d \tau \ee
being the nucleon total proper time.

\begin{figure}[t!]
\label{fin}
\resizebox{9.1cm}{!}{\includegraphics{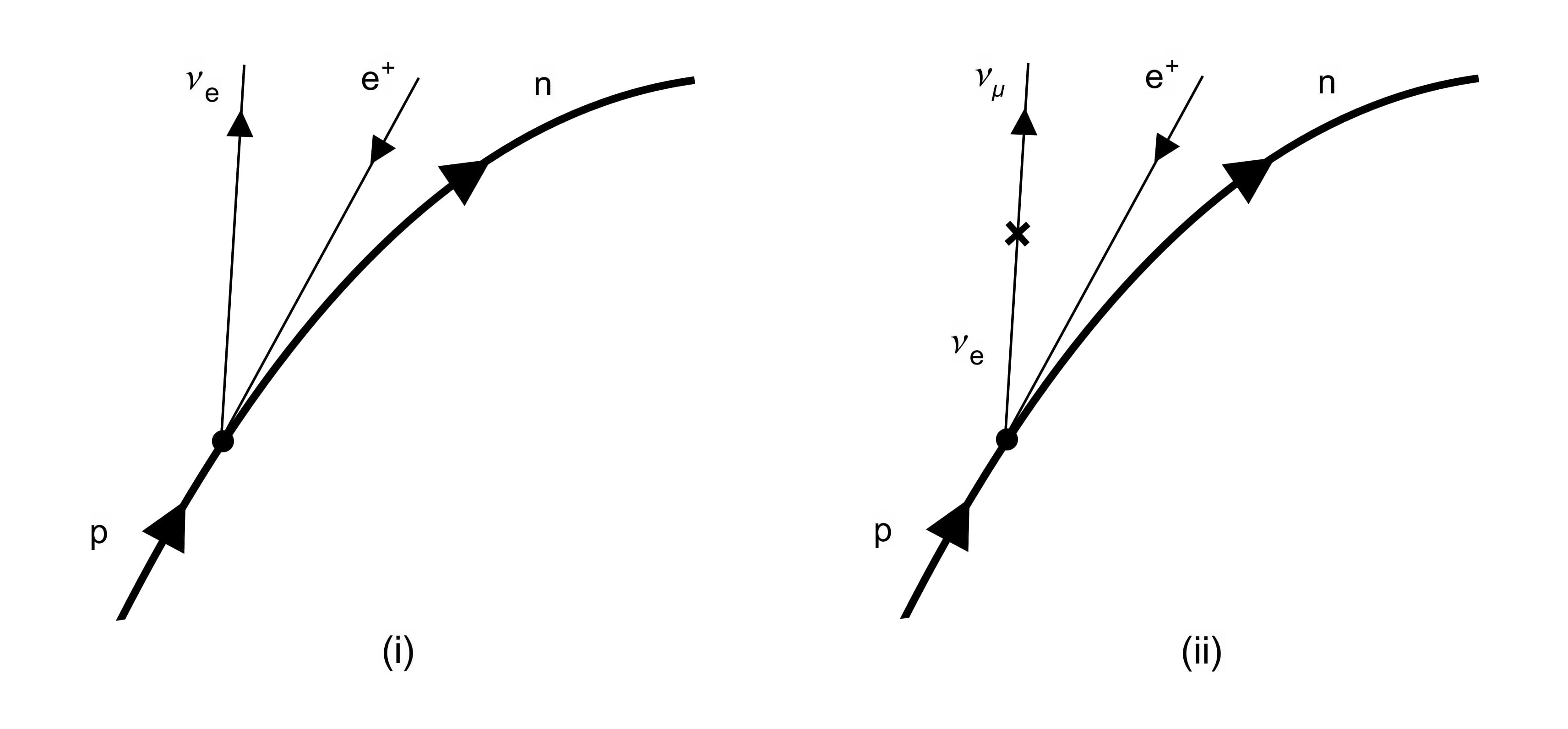}}
\centering
\caption{Decay process in the inertial frame without (i) 
and with (ii) neutrino oscillation. Time flows in the vertical direction.}
\end{figure}

In Ref.~\cite{Blasone:2018czm}, it has been shown that the transition rate for the process Eq.~(\ref{drt}) is
\be
\label{eqn:inertresultat}
\Gamma^{(\nu_e)}_{in}\ =\ \cos^4\theta\, \Gamma_{1}\,+\,\sin^4\theta\,\Gamma_{2}\,+\,\cos^2\theta\sin^2\theta\,\Gamma_{12}.
\ee
Here, we have introduced the shorthand notation
\begin{equation}
\label{integral}
\Gamma_{j}\ \equiv\ \frac{1}{T}\sum_{\sigma_\nu,\sigma_e}\frac{G_F^2}{2^8\pi^6}\int d^3k_\nu\int d^3k_e\,{\big|\mathcal{I}_{\sigma_\nu\sigma_e}(\omega_{\nu_j},\omega_e)\big|}^2,
\end{equation}
where $j=1,2$ and 
\bea
\label{integral12}
\Gamma_{12}\ &\equiv& \frac{1}{T}\sum_{\sigma_\nu,\sigma_e}\frac{G_F^2}{2^8\pi^6}\int d^3k_\nu\int d^3k_e\\[2mm]\nonumber
&&\times\Big[\mathcal{I}_{\sigma_\nu\sigma_e}(\omega_{\nu_1},\omega_e)\,\mathcal{I}^{\hspace{0.2mm}*}_{\sigma_\nu\sigma_e}(\omega_{\nu_2},\omega_e)\,+\,\mathrm{c.c.}\Big].
\eea
The aim of the calculations contained in Ref.~\cite{Blasone:2018czm} is to exhibit the equality between Eq.~(\ref{eqn:inertresultat}) and its counterpart in the accelerated frame (see below), which guarantees the validity of the principle of general covariance.

\medskip

At this point, it must be emphasized that in the above calculation an infinite proper time interval is considered, which allows for the electron neutrino produced in the proton decay to oscillate. Thus, we should take into account not only the process contemplated in Eq.~(\ref{i}), but also the following one:
\be\label{prm}
\mathrm{(ii)}\quad p\,\to\, n\, +\, e^{+}\,+\,\nu_\mu\,.
\ee
The above relation must be intended in the sense of Fig.~1: although it is true that  the lepton charge must necessarily be conserved in the vertex
(at tree-level),  as soon as the outgoing neutrino is produced, flavor oscillations will inevitably take place. We remark that this process has not been included in the analysis of Ref.~\cite{Blasone:2018czm}, without however affecting the validity of the results there contained.

The transition amplitude for the process (ii) is indeed non-vanishing and is given by
\bea\label{trampm}
&\mathcal{A}^{(\nu_\mu)}_{\mathrm{(ii)}}&=\langle n|\otimes\langle e^{+},\nu_{\mu}|\hat{S}_{I}|0\rangle\otimes|p\rangle\\[2mm]\nonumber
&=&\hspace{-3.5mm}-\frac{G_F}{2^4\pi^3}\cos\theta\sin\theta\Big[\mathcal{I}_{\sigma_\nu\sigma_e}(\omega_{\nu_1},\omega_e)-\mathcal{I}_{\sigma_\nu\sigma_e}(\omega_{\nu_2},\omega_e)\Big].
\eea
In terms of $\Gamma$, the quantity ${\cal A}_{\mathrm{(ii)}}^{(\nu_{\mu})}$ of Eq.~(\ref{trampm}) associated to the process (ii) leads to the following transition rate:
\be
\label{eqn:inertresultat2}
\Gamma^{(\nu_\mu)}_{in}=\cos^2\theta\sin^2\theta\lf(\Gamma_1+\Gamma_2-\Gamma_{12}\ri)\,,
\ee
where all the contributions in the r.h.s. of the above expression have already been introduced in Eqs.~(\ref{integral}) and (\ref{integral12}). We notice that the above transition rate is proportional to $\sin^22\theta$, thus showing that it is originated by interference.

Finally, observe that
\be\label{ins}
\Gamma_{in}\equiv\Gamma^{(\nu_e)}_{in}+\Gamma^{(\nu_\mu)}_{in}=\cos^2\theta\,\Gamma_1+\sin^2\theta\,\Gamma_2\,.
\ee

\medskip

%%%%%%%%%%%%%%%%%%%%%%%%%%%%%%%%%%%%%%%%%%%%%%%%%%%%%%%%%%%%%%%%%%%
{\em Accelerated frame}. From the point of view of an observer comoving with the proton, the only possible way to render the particle's decay viable is to suppose the existence of a thermal bath of electrons, neutrinos and the corresponding antiparticles~\cite{Blasone:2018czm}. In such conditions, recalling that here we consider the electron decay channel only, three processes can be taken into account to match the inertial process Eq.~(\ref{i}):
\begin{subequations}\label{threeprocesses}
\begin{align}
\label{iii}
\mathrm{(iii)}\quad p^{+}\,+\,e^{-}\,\rightarrow\, n\,+\,\nu_e,\\[1.5mm]
\label{iv}
\mathrm{(iv)}\quad p^{+}\,+\,\overline{\nu}_e\,\rightarrow\, n\,+\,e^{+},\\[1.5mm]
\label{v}
\mathrm{(v)}\quad p^{+}\,+\,e^{-}\,+\,\overline{\nu}_e\,\rightarrow\, n.
\end{align}
\end{subequations}
In order to calculate the proton's decay rate in the accelerated frame, we need the expansion for the fermion fields in Rindler coordinates, that is
\bea
\nonumber
\hspace{-3mm}\hat{\Psi}&=& \sum_{\sigma=\pm}\int_{0}^{+\infty}\hspace{-3mm}d\omega\int\frac{d^2k}{{(2\pi)}^{\frac{3}{2}}}\left[{e^{i(-\omega v/a\,+\,k_\alpha x^\alpha)}}\hspace{0.2mm}u_{\sigma}^{(\omega)}\hat{b}_{\textbf{w}\sigma}\hspace{0.2mm}\right.\\[2mm]\label{Rindexp}
&&+\left.{e^{i(\omega v/a\,+\,k_\alpha x^\alpha)}}\hspace{0.2mm}u_{-\sigma}^{(-\omega)}\hat{d}_{\textbf{w}\sigma}^{\dagger}\hspace{0.2mm}\right],
\eea
where $k_\al x^\al=k_xx+k_yy$, $\omega$ is the Rindler frequency which does not satisfy any dispersion relation,  $\textbf{w}=(\omega,k^x,k^y)$ and $u^{(\omega)}_\sigma$ is defined as in Ref.~\cite{Blasone:2018czm}.

By way of illustration, let us consider the scattering (iii) of Eq.~(\ref{iii}) (see also Fig.~2); similar calculations can be carried out for the processes  in Eqs.~(\ref{iv}), (\ref{v}). We obtain~\cite{Blasone:2018czm}
\bea
\label{first}
&&\mathcal{A}^{(\nu_e)}_{\mathrm{(iii)}}\equiv \left\langle n\right|\otimes\langle\nu_{e}|\hspace{0.2mm}\hat{S}_{I}\hspace{0.2mm}|e^{-}\rangle\otimes\left|p\right\rangle\\[2mm]\nonumber
&&=\frac{G_F}{(2\pi)^2}\lf[\cos^2\theta\,\mathcal{J}^{(1)}_{\sigma_\nu\sigma_e}(\omega_\nu, \omega_e)+\sin^2\theta\,\mathcal{J}^{(2)}_{\sigma_\nu\sigma_e}(\omega_\nu, \omega_e)\ri],
\eea
where we have assumed equal frequencies  and polarizations   for different neutrinos with definite masses, whereas the quantity denoted with $\mathcal{J}$ is equal to
\be
\label{eqn:j}
\mathcal{J}^{(i)}_{\sigma_\nu\sigma_e}(\omega_\nu, \omega_e)\ =\ \delta\big(\omega_e-\omega_{\nu}-\Delta m\big)\,\bar{u}_{\sigma_{\nu}}^{(i,\omega_{\nu})}\gamma^0 u_{\sigma_{e}}^{(\omega_{e})}.
\ee
with $i=1,2$. The spinor component related to the neutrino field contains the information on the mass of $\nu_i$, and by means of the current hypothesis this is the only difference between the functions ${\cal J}^{(i)}$ with different indexes.

\begin{figure*}[t!]\label{fig2}
\resizebox{18.5cm}{!}{\includegraphics{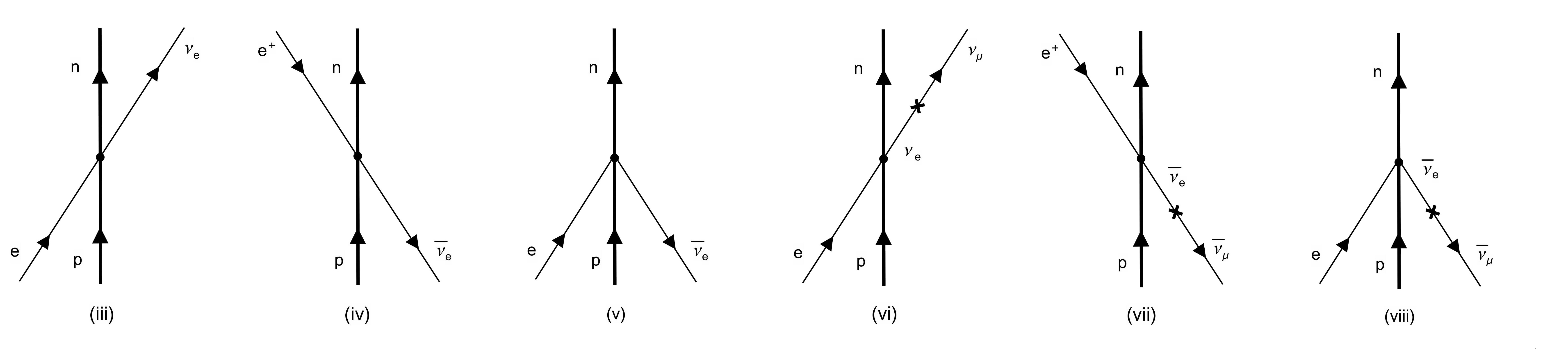}}
\centering
\caption{Decay processes in the accelerated frame. 
Oscillations of neutrinos in the Unruh thermal bath are considered in the last two diagrams. Time flows in the vertical direction.}
\end{figure*}

The sum of the transition rates for the three processes in 
Eqs.~(\ref{threeprocesses}) yields~\cite{Blasone:2018czm}
\begin{equation}
\label{gammacc}
\Gamma_{acc}^{(\nu_e)}\ =\ %\frac{4G_{F}^{2}m_{e}}{\pi^3 a^{2}e^{\pi\Delta m/a}}\int_{-\infty}^{+\infty}\hspace{-1.5mm}d\omega_e\left[m_{1}\cos^{4}\theta\,R_{1}^{2}(\omega_e)+m_{2}\sin^{4}\theta\,R_{2}^{2}(\omega_e)+2\cos^{2}\theta\sin^{2}\theta\sqrt{m_{1}m_{2}}R_{1}(\omega_e)R_{2}(\omega_e)\right]\\[2mm]
\cos^4\theta\, \widetilde\Gamma_{1}\,+\,\sin^4\theta\,\widetilde\Gamma_{2}\,+\,\cos^2\theta\sin^2\theta\,\widetilde\Gamma_{12},
\end{equation}
where
\begin{equation}
\label{integralbis}
\widetilde\Gamma_{j}\ \equiv\ \mathcal{N}\int_{-\infty}^{+\infty}\hspace{-1.5mm}d\omega\hspace{0.4mm}\mathcal{R}_{j}(\omega),\qquad j\,=\,1,2,
\end{equation}
and $\mathcal{N}={2\pi^{-7}a^{-2}G_{F}^{2}}{e^{-\pi\Delta m/a}}$. 
The functions $\widetilde\Gamma_{12}$ and $\mathcal{R}_{j}$ in Eqs.~\eqref{gammacc}
and~\eqref{integralbis}, respectively, are defined by
\begin{widetext}
\bea
&&{}\hspace{-7mm}\mathcal{R}_j(\omega)\!=\!\!\int \!d^2 k_{\nu}d^2k_{e}\,l_{\nu_j}\,l_{e}\Bigl|K_{\frac{i}{a}\tilde{\omega} +\frac{1}{2}}\left(\frac{l_{\nu_j}}{a}\right) K_{\frac{i}{a}\omega+\frac{1}{2}}\left(\frac{l_{e}}{a}\right)\Bigr|^2\!\!+ m_{\nu_j}m_e\mathrm{Re}\left[\!\int \!d^2k_{\nu} d^2k_eK^2_{\frac{i}{a}\tilde{\omega}-\frac{1}{2}}\left(\frac{l_{\nu_j}}{a}\right) K^2_{\frac{i}{a}\omega+\frac{1}{2}}\left(\frac{l_{e}}{a}\right)\right],
\label{R}
\eea
and
\begin{eqnarray}\nonumber
\label{int1212}
\widetilde\Gamma_{12}&=&\frac{\mathcal{N}}{\sqrt{l_{\nu_1}l_{\nu_2}}}\int d\omega d^2k_e d^2k_\nu\Biggl\{\, l_e\Bigl|K_{\frac{i}{a}\omega+\frac{1}{2}}\left(\frac{l_e}{a}\right)\Bigr|^2\big(\kappa_\nu^2+m_{\nu_1}m_{\nu_2}+l_{\nu_1}l_{\nu_2}\big)\mathrm{Re}\left[K_{\frac{i}{a}\tilde{\omega}+\frac{1}{2}}\left(\frac{l_{\nu_1}}{a}\right)K_{\frac{i}{a}\tilde{\omega} -\frac{1}{2}}\left(\frac{l_{\nu_2}}{a}\right)\right]
\\[2mm]
&&+\,m_e  \big(l_{\nu_1}m_{\nu_2}+l_{\nu_2}m_{\nu_1}\big) \,\mathrm{Re}\left[K^2_{\frac{i}{a}\omega+\frac{1}{2}}\left(\frac{l_e}{a}\right)K_{\frac{i}{a}\tilde{\omega} -\frac{1}{2}}\left(\frac{l_{\nu_1}}{a}\right) K_{\frac{i}{a}\tilde{\omega} -\frac{1}{2}}\left(\frac{l_{\nu_2}}{a}\right)\right]\Biggr\},
\end{eqnarray}
\end{widetext}
where $l_{\nu_j(e)}\equiv\sqrt{m^2_{\nu_j(e)}+{(k^x)}^2+{(k^y)}^2}$ and we 
have used the shorthand notation $\tilde{\omega}\equiv\omega-\Delta m$.

In Ref.~\cite{Blasone:2018czm} it was shown that, by taking
into account the three interactions in Eq.~\eqref{threeprocesses} of the proton with the particles of the thermal bath, the quantity $\Gamma^{(\nu_e)}_{acc}$ matches the corresponding decay rate $\Gamma^{(\nu_e)}_{in}$ evaluated in the inertial frame. In particular, it is possible to exhibit~\cite{Blasone:2018czm} that $\Gamma_i=\widetilde{\Gamma}_i$ for $i=1,2$, whereas $\Gamma_{12}$ and $\widetilde{\Gamma}_{12}$ are equal to each other only up to a first-order expansion in the parameter $\delta m\equiv m_{\nu_2}-m_{\nu_1}$. In such an approximation, Pontecorvo  states Eq.~\eqref{Ponte} can be identified with the exact flavor neutrino states, defined as  eigenstates of flavor charges \cite{Blasone:1995zc,Blasone:2001qa}.

On the other hand, we have seen above that an additional contribution to the proton decay rate has to be considered, which in the inertial frame is represented by the process (ii) in Fig.~1. Guided by the principle of general covariance, we now seek the corresponding processes in the comoving frame which should lead to  the same result. 

To this aim, we consider the three following contributions as potential candidates for the non-inertial counterpart of the decay Eq.~(\ref{prm}):
\begin{subequations}\label{threeprocessesmu}
\begin{align}
\label{vi}
\mathrm{(vi)}\quad p^{+}\,+\,e^{-}\,\rightarrow\, n\,+\,\nu_\mu,\\[1.5mm]
\label{vii}
\mathrm{(vii)}\quad p^{+}\,+\,\overline{\nu}_\mu\,\rightarrow\, n\,+\,e^{+},\\[1.5mm]
\label{viii}
\mathrm{(viii)}\quad p^{+}\,+\,e^{-}\,+\,\overline{\nu}_\mu\,\rightarrow\, n,
\end{align}
\end{subequations}
which are depicted in Fig.~2 in the last three diagrams. Note that, while the process (vi) is of the same type of (ii) since it entails an oscillation of the emitted (electron) neutrino, the processes (vii) and (viii) are essentially due to the oscillation of an (muon) antineutrino that is already present in the Unruh thermal bath.

In order to legitimate the validity of our assumption, we need to perform the same calculations that lead to the decay rate of Eq.~(\ref{gammacc}). The outcome of this procedure turns out to be
\be\label{muacc}
\Gamma^{(\nu_\mu)}_{acc}=\cos^2\theta\,\sin^2\theta\left(\widetilde{\Gamma}_1+\widetilde{\Gamma}_2-\widetilde{\Gamma}_{12}\right)\,.
\ee
By virtue of the aforesaid observations contained in detail in Ref.~\cite{Blasone:2018czm} which allow us to state that $\Gamma_{in}^{(\nu_e)}=\Gamma_{acc}^{(\nu_e)}$, it is possible to infer that such an equivalence holds also between the decay rates of Eqs.~(\ref{eqn:inertresultat2}) and (\ref{muacc}).

Moreover, if we compute the total comoving decay rate which includes neutrino oscillations, we deduce that
\be\label{mutot}
\Gamma_{acc}=\Gamma^{(\nu_e)}_{acc}+\Gamma^{(\nu_\mu)}_{acc}=\cos^2\theta\,\widetilde{\Gamma}_1+\sin^2\theta\,\widetilde{\Gamma}_2\,.
\ee
By comparing this with the total inertial decay rate of Eq.~(\ref{ins}), we find that 
\be\label{fr}
\Gamma_{in}=\Gamma_{acc}\,,
\ee
which means that such a result does not depend on the quantities $\Gamma_{12}$ and $\widetilde{\Gamma}_{12}$, whose treatment would require additional computational effort~\cite{Blasone:2018czm}. 

Remarkably, Eq.~(\ref{fr}) not only involves a generalization of the analysis of the accelerated proton decay to the case in which the produced neutrino oscillates, but it also unambiguously corroborates our guess of selecting the processes in Eqs.~(\ref{threeprocessesmu}) as the counterpart for the decay (ii) in the inertial frame. Hence, the requirement of the principle of general covariance clearly results in the necessity of having an Unruh thermal bath containing flavor neutrinos which do oscillate. 

\medskip

\emph{Conclusions.} In this Letter, we have  extended the formalism firstly introduced in Refs.~\cite{Matsas:1999jx} and then developed in Refs.~\cite{Aluw,Blasone:2018czm,Cozzella:2018qew}  to the case in which
   neutrino oscillations are  taken into account.

In these works, the principle of general covariance has been exploited in the analysis of accelerated proton decay to probe the internal consistency of the theory and to shed light on new aspects of this problem.
In such a perspective, here we have further studied this subject by considering flavor oscillations for the emitted neutrinos. By enforcing general covariance, we have then found that the Unruh radiation ``seen'' by the accelerated proton must involve oscillating neutrinos. This is a novel feature which emerges in a natural way in the present approach. 

A further interesting observation that can be deduced from our analysis is related to the identities Eqs.~(\ref{ins}) and (\ref{mutot}) that are true for the inertial and the comoving frame, respectively. For this purpose, we recall that the decay rates appearing in the aforementioned equations have been computed by employing neutrino flavor states as asymptotic states. However, we note that similar relations also hold  for the quantities $\Gamma^{(\nu_1)}$ and $\Gamma^{(\nu_2)}$ calculated in Ref.~\cite{Cozzella:2018qew} using neutrino mass eigenstates as fundamental objects. We then have
\be\label{ff}
\Gamma^{(\nu_1)}+\Gamma^{(\nu_2)}=\Gamma^{(\nu_e)}+\Gamma^{(\nu_\mu)}\,,
\ee
where the equality has to be regarded both in the inertial and the comoving frames. Such an equation constitutes a consistency check for the correctness of the calculations in Refs.~\cite{Blasone:2018czm} and \cite{Cozzella:2018qew}.
The physical meaning of Eq.~\eqref{ff} can be understood by considering the charges for mixed neutrino fields as derived from Noether's theorem~\cite{Blasone:2001qa}. 
Indeed, by denoting with
\be\nonumber
Q_i=\int d^3x\, \Psi_{\nu_i}^\dag(x)\Psi_{\nu_i}(x)\,, \quad i=1,2,
\ee
the conserved charges for the neutrino fields with definite masses and with 
\be\non
Q_\al(t)=\int d^3x\,\Psi_{\nu_\al}^\dag(x)\Psi_{\nu_\al}(x)\,, \quad \al=e,\mu,
\ee 
the (time-dependent) flavor charges, one can see that~\cite{Blasone:2001qa}
\be\non
Q=\sum_i Q_i = \sum_\al Q_\al(t)\,,
\ee
where $Q$ represents the total charge.
The above relation can be interpreted as the conservation of the total lepton number. On the one hand, this can be viewed as the sum of two separately conserved family lepton numbers, when no mixing is present; on the other hand, the same conserved number is obtained by the sum of non-conserved flavor charges, which indeed are associated to neutrino oscillations.

Apart from its relevance in the context
of neutrino mixing and oscillations, we stress
that the Unruh effect is an excellent
benchmark for both testing well-established predictions 
and pointing out novel effects in 
fundamental physics, as it combines such wide domains as
general relativity, quantum mechanics and thermodynamics. 
For instance, in Refs.~\cite{nonthermal} it has been shown
that the Unruh spectrum may exhibit 
exotic non-thermal corrections
even within the standard QFT, thus emphasizing how 
such a framework provides an active forge
of still unexplored scenarios. 
A similar non-thermal behavior 
has been obtained in Refs.~\cite{Luciano:2019mrz},
where Planck scale effects on the Unruh bath
have been derived in the context
of the Generalized Uncertainty Principle~\cite{gup}.
Further features of the Unruh effect
may be addressed in connection with entanglement 
properties for accelerated observers, 
whose implications have been investigated also in the context of black hole physics~\cite{Hwang:2010ib}.
In particular, in Ref.~\cite{FuentesSchuller:2004xp}, it has been
proved that entanglement turns out to be an 
observer-dependent quantity in non-inertial frames, due to
the Unruh radiation. The question thus arises
as to how this setting is modified 
in the presence of mixed neutrinos, particularly in view
of the discussion of Refs.~\cite{blasent}.

As a final remark, we stress that in the current Letter we have made use of the simplest framework of neutrino mixing among two generations. The extension to three flavors is in principle straightforward, and represents one of the future directions of our investigation. We envisage that the presence of CP violation may introduce interesting additional features which would enrich the non-trivial structure of the Unruh radiation.

\end{document}